\documentclass{article}
\usepackage{mathrsfs}
\usepackage{amsmath}
\usepackage{amsfonts}
\usepackage{graphicx,color}
\usepackage{feynmp}
\usepackage{psfrag}
\usepackage{cite}
\usepackage{epsfig}
\begin{document}
\newcommand{\bra}[1]{\langle#1|}
\newcommand{\ket}[1]{|#1\rangle}

\begin{titlepage}
\begin{flushright}
ZU-TH 13/07\\
\end{flushright}
\begin{center}
\vskip 3.5cm
{\Large \bf Constraints On Unparticle Physics In Electroweak Gauge Boson Scattering}
\vskip 1cm
{\large N.Greiner}
\vskip .7cm 
{\it Institut f\"ur Theoretische Physik, Universit\"at Z\"urich,
Winterthurerstrasse 190, CH-8057 Z\"urich, Switzerland} 
\vskip 2.6cm 

\begin{abstract} 
The existence of scale invariant physics would lead to new phenomena in particle physics that could be detected
at the LHC. In this paper we exploit the effects of these unparticles in $WW \to WW$ scattering. From the requirement
of unitarity we derive constraints on unparticle physics. We show that the existence of unparticles would lead to deviations
in differential cross sections which can be measured. These deviations are sensitive on the scale dimension and on
the spin characteristics of the unparticles.

\end{abstract}
\end{center}
\end{titlepage}

\Large \bfseries 1 \hspace{0.5cm} Introduction\\ \\
\mdseries
Recently H.Georgi proposed that there might be a sector of scale invariant physics that decouples at a large scale \cite{Georgi:2007ek}, \cite{Georgi:2007si}. At high energies this theory contains both the fields of the Standard Model and the fields of the scale invariant physics. This invariant sector is called BZ fields \cite{Banks:1981nn}. The two types of fields interact via exchange of particles with a mass scale $M_{\cal U}$. Renormalization of the couplings of the BZ fields causes dimensional transmutation at the scale $\Lambda_{\cal U}$ as scale invariance emerges. Below that scale $\Lambda_{\cal U}$ one ends up with a nonrenormalizable effective theory. If there is physics that is scale invariant the low energy sector can be associated with "unparticles". These unparticles behave like invisible massless particles and are described by unparticle operators $O_{\cal U}$. As it is not a priori clear what lorentz structure these unparticles have, one can construct scalar-, vector-, and tensor-operators $O_{\cal U}$,$O^{\mu}_{\cal U}$,$O^{\mu\nu}_{\cal U}$. As suggested in \cite{Georgi:2007si} one for example has\\
\begin{equation}\label{op} \frac{\lambda_1}{\Lambda^{d_u}}F_{\alpha\beta}F^{\alpha\beta}O_{\cal U}, \quad \frac{\lambda_2}{\Lambda^{d_u}}F_{\mu\alpha}F^{\alpha}_{\nu}O_{\cal U}^{\mu\nu}, \end{equation}
where $F_{\alpha\beta}$ is the field strength tensor of any bosonic field of the Standard model and $\frac{\lambda_i}{\Lambda^{d_u}}$ denotes the effective coupling constant. $d_{\cal U}$ is the scaling dimension of the unparticle operators $O_{\cal U}$.\\
Following the notation of \cite{Georgi:2007si}, the propagator of a vector-like unparticle is given by\\
\begin{flalign}\begin{split} \int e^{iPx}\bra{0}T(O^{\mu}_{\cal U}(x)O^{\nu}_{\cal U}(0)\ket{0}d^4x \qquad\\
=i\frac{A_{d_{\cal U}}}{2} \int_{0}^{\infty} (M^2)^{d_{\cal U}-2}\frac{-g^{\mu\nu}+P^{\mu}P^{\nu}/P^2}{P^2-M^2+i\epsilon}dM^2 \\ =i\frac{A_{d_{\cal U}}}{2} \frac{-g^{\mu\nu}+P^{\mu}P^{\nu}/P^2}{\sin(d_{\cal U})\pi}(-P^2-i\epsilon)^{d_{\cal U}-2},\end{split}\end{flalign}\\ \\
with\\
\begin{equation} A_{d_{\cal U}}=\frac{16\pi^{5/2}}{(2\pi)^{2d_{\cal U}}}\frac{\Gamma(d_{\cal U}+1/2)}{\Gamma(d_{\cal U}-1)\Gamma(2d_{\cal U})}. \end{equation} \\ \\ \\
The propagator for a scalar unparticle is then given by\\
\begin {equation} i\frac{A_{d_{\cal U}}}{2} \frac{(-P^2-i\epsilon)^{d_{\cal U}-2}}{\sin(d_{\cal U}\pi)}. \end{equation} \\
This was also derived independently in \cite{Cheung:2007ue}.The scale dimension of the unparticle operator is constrained to $1<d_{\cal U}<2$. \\
So far there have been several publications that exploited phenomenological aspects of unparticles \cite{Li:2007by,Lu:2007mx,Duraisamy:2007aw,Aliev:2007qw,Ding:2007bm,Luo:2007bq,Cheung:2007ue,Liao:2007bx,Chen:2007vv,Lu:2007mx,Stephanov:2007ry,Fox:2007sy}, but no attempts have been made in the purely bosonic sector.\\
Effective theories described by such operators as in (\ref{op}) cannot be valid to arbitrary energies. Their applicability is bound by the requirement of unitarity. That means that we can constrain the strength of the coupling constant if we demand the theory to be applicable to a certain energy which of course has to be well below the energy scale $\Lambda_{\cal U}$.\\
Considering the bosonic sector, it has been observed in the past, that severe constraints on the somehow related dimension six operators come from the $WW \to WW$ scattering when all $W$-bosons are longitudinally polarized \cite{Gounaris:1994cm}. Because of that we are now also going to focus on the $WW$ scattering including unparticles.\\ \\
The paper is organised as follows. We will first review some important aspects on $WW \to WW$ scattering in the Standard Model and then include the exchange of unparticles. By demanding that unitarity must not be violated we calculate upper bounds on the strength of the coupling constant. Looking then at the angular distribution of the outgoing $W$-bosons we find characteristic dependencies on the scale dimension $d_{\cal U}$ that allows to discriminate an exchange of an unparticle from the Standard Model scenario and may even allow to discriminate a scalar unparticle from a vector unparticle.\\ \\ \\ \\ \\ 
\Large \bfseries 2 \hspace{0.5cm} WW $\to$ WW scattering\\ \\
\mdseries
As figure 1 shows there are seven diagrams that contribute to the $WW \to WW$ scattering in the Standard Model. Let $k$ be the momentum of the incoming $W$-bosons in the c.m. frame. Looking at the case when all $W$-bosons are longitudinally polarised each diagram has a leading contribution to the amplitude that is proportional to $k^4$,  except the two Higgs diagrams. As we are interested in the behaviour at high energies, i.e. in regions of around 1 TeV, it is sufficient to only look at the term with the highest order in $k$. All other terms can be omitted. If one adds up these five diagrams, the $k^4$ terms cancel and the amplitude is proportional to $k^2$. The two Higgs exchange diagrams are also proportional to $k^2$ and these two contributions cancel so that the leading term of the whole amplitude is a constant in $k$ as one takes the high energy limit. So we have no violation of unitarity.\\ \\ 
\begin{figure}
\begin{center}
\begin{fmffile}{wwtoww}
\parbox{3cm}{\begin{fmfgraph*}(100,60) \fmfleft{i1,i2} \fmfright{o1,o2} \fmf{photon}{i1,v1} \fmf{photon}{i2,v1} \fmfv{decor.shape=circle,decor.size=1}{v1} \fmf{photon,label=$\gamma$/Z}{v1,v2} \fmfv{decor.shape=circle,decor.size=1}{v2} \fmf{photon}{v2,o1} \fmf{photon}{v2,o2} \fmflabel{$W$}{i1} \fmflabel{$W$}{i2} \fmflabel{$W$}{o1} \fmflabel{$W$}{o2}
\end{fmfgraph*}}
\hspace{1.5cm}
\parbox{3cm}{\begin{fmfgraph*}(100,60) \fmfleft{i1,i2} \fmfright{o1,o2} \fmf{photon}{i1,v1} \fmf{photon}{i2,v1} \fmfv{decor.shape=circle,decor.size=1}{v1} \fmf{scalar,label=H}{v1,v2} \fmfv{decor.shape=circle,decor.size=1}{v2} \fmf{photon}{v2,o1} \fmf{photon}{v2,o2} \fmflabel{$W$}{i1} \fmflabel{$W$}{i2} \fmflabel{$W$}{o1} \fmflabel{$W$}{o2}
\end{fmfgraph*}}
\hspace{1.5cm}
\parbox{3cm}{\begin{fmfgraph*}(100,60) \fmfleft{i1,i2} \fmfright{o1,o2} \fmf{photon}{i1,v1} \fmfv{decor.shape=circle,decor.size=1}{v1} \fmf{photon}{i2,v2} \fmfv{decor.shape=circle,decor.size=1}{v2}  \fmf{photon,label=$\gamma$/Z}{v1,v2} \fmf{photon}{v2,o2} \fmf{photon}{v1,o1} \fmflabel{$W$}{i1} \fmflabel{$W$}{i2} \fmflabel{$W$}{o1} \fmflabel{$W$}{o2} \end{fmfgraph*}  }\\
\addvspace{1.5cm}
\parbox{3cm}{\begin{fmfgraph*}(100,60) \fmfleft{i1,i2} \fmfright{o1,o2} \fmf{photon}{i1,v1} \fmfv{decor.shape=circle,decor.size=1}{v1} \fmf{photon}{i2,v2} \fmfv{decor.shape=circle,decor.size=1}{v2}  \fmf{scalar,label=H}{v2,v1} \fmf{photon}{v2,o2} \fmf{photon}{v1,o1} \fmflabel{$W$}{i1} \fmflabel{$W$}{i2} \fmflabel{$W$}{o1} \fmflabel{$W$}{o2} \end{fmfgraph*}  }
\hspace{1.5cm}
\parbox{3cm}{\begin{fmfgraph*}(100,60) \fmfleft{i1,i2} \fmfright{o1,o2} \fmf{photon}{i1,v1} \fmfv{decor.shape=circle,decor.size=1}{v1} \fmf{photon}{i2,v1} \fmf{photon}{v1,o1} \fmf{photon}{v1,o2} \fmflabel{$W$}{i1} \fmflabel{$W$}{i2} \fmflabel{$W$}{o1} \fmflabel{$W$}{o2}\end{fmfgraph*}}
\end{fmffile}\\ \addvspace{1cm}
\caption{WW $\rightarrow$ WW scattering. }
\end{center}
\end{figure}
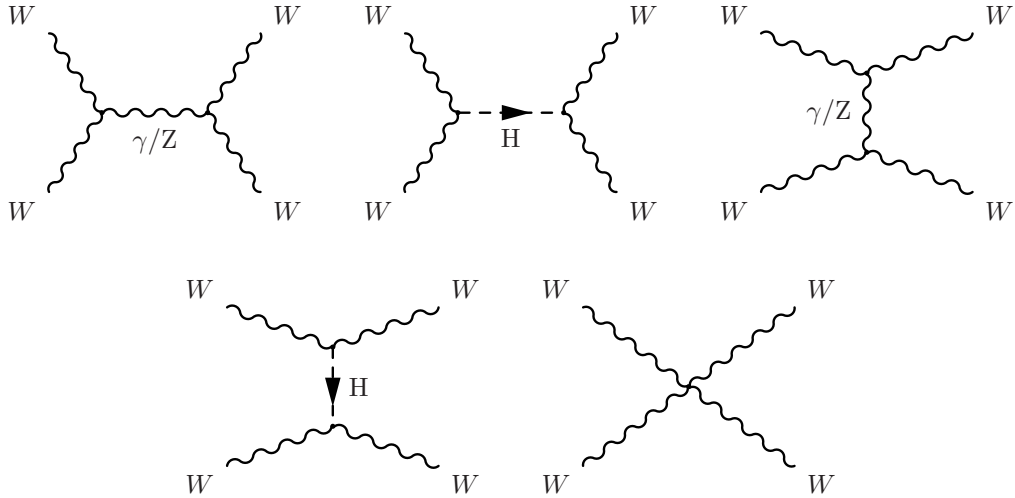
Independent whether we are considering scalar unparticles or vector-like unparticles we have two contributions, namely the exchange of an unparticle in the s-channel and in the t-channel. This is illustrated in figure 2. We now have additional contributions which are not cancelled and that will cause unitarity violation if we go to sufficient high energies.\\ \\
We start with considering a scalar unparticle. The effective operator takes the form\\
\begin{equation*} \frac{\lambda_1}{\Lambda^{d_{\cal U}}}\vec{W}_{\mu\nu}\vec{W}^{\mu\nu}O_{\cal U}. \end{equation*}
We therefore get new Feynman rules for the coupling between the unparticles and the $W$-bosons due to the structure of the given operators. Let $\mu$ be the lorentz index of the $W^+$ with momentum $q_+$ and $\nu$ the lorentz index of the $W^{-}$ with momentum $q_{-}$. Then for a scalar unparticle the vertex factor is\\
\begin{figure}
\begin{center}
\begin{fmffile}{unparticles}
\parbox{3cm}{\begin{fmfgraph*}(100,60) \fmfleft{i1,i2} \fmfright{o1,o2} \fmf{photon}{i1,v1} \fmf{photon}{i2,v1} \fmfv{decor.shape=circle,decor.size=1}{v1} \fmf{scalar,label=\cal U}{v1,v2} \fmfv{decor.shape=circle,decor.size=1}{v2} \fmf{photon}{v2,o1} \fmf{photon}{v2,o2} \fmflabel{$W$}{i1} \fmflabel{$W$}{i2} \fmflabel{$W$}{o1} \fmflabel{$W$}{o2}
\end{fmfgraph*}}
\hspace{1.5cm}
\parbox{3cm}{\begin{fmfgraph*}(100,60) \fmfleft{i1,i2} \fmfright{o1,o2} \fmf{photon}{i1,v1} \fmfv{decor.shape=circle,decor.size=1}{v1} \fmf{photon}{i2,v2} \fmfv{decor.shape=circle,decor.size=1}{v2}  \fmf{scalar,label=\cal U}{v2,v1} \fmf{photon}{v2,o2} \fmf{photon}{v1,o1} \fmflabel{$W$}{i1} \fmflabel{$W$}{i2} \fmflabel{$W$}{o1} \fmflabel{$W$}{o2} \end{fmfgraph*}  }
\end{fmffile} \\ \addvspace{1cm}
\caption{Contribution of unparticles to WW $\rightarrow$ WW scattering.}
\end{center}
\end{figure}
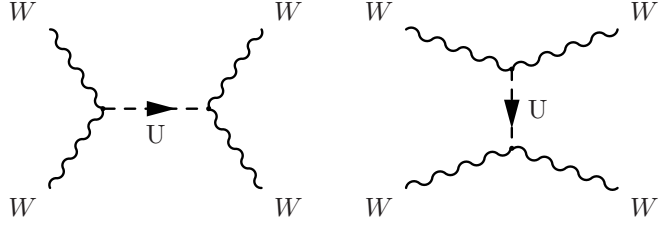
\begin {equation*} 4i\frac{\lambda_1}{\Lambda^{d_{\cal U}}}(q_{+}^{\nu}q_{-}^{\mu}-g^{\mu\nu}(q_{+}\cdot q_{-})). \end{equation*}
Naively one would expect that the leading contribution comes from the state when all $W$-bosons are longitudinally polarised as it is for instance for dimension six operators. But due to the gauge structure of this operator, this is not the case. Explicit calculation shows, that the amplitude with four longitudinal bosons only grows with $k^{2d_{\cal U}-4}$ whereas the leading term is $\sim k^{2d_{\cal U}}$ and solely comes from the amplitudes where all bosons are transversely polarised. That the leading term is $\sim k^{2d_{\cal U}}$ is not surprising as we expect the same behaviour as an exchange of a Higgs boson when $d_{\cal U}=1$.\\
For a vector-like unparticle the high energy behaviour depends on the structure of the operater $O_{\cal U}^{\mu\nu}$. We assume a gauge-invariant form and choose it as $O_{\cal U}^{\mu\nu}=\partial^{\mu}O_{\cal U}^{\nu}-\partial^{\nu}O_{\cal U}^{\mu}$. Because we made a certain assumption on the form of the unparticle operator $O_{\cal U}^{\mu\nu}$ we have to mildly adapt the coupling constant to get the right dimensions. We therefore change the coupling constant to $\frac{\lambda_2}{\Lambda^{d_{{\cal U}}+1}}$. Off course it is not at all clear that this is the whole operator, one can think of adding a non-abelian part but these parts would not contribute to our diagrams. Our effective operator then has the form\\
\begin{equation*} \frac{\lambda_2}{\Lambda^{d_u+1}}\vec{W}_{\mu\alpha}\vec{W}^{\alpha}_{\nu}O_{\cal U}^{\mu\nu}. \end{equation*}
From that expression we derive the following Feynman rule:\\
\begin{flalign*} 2\frac{\lambda_2}{\Lambda^{d_{\cal U}+1}}(q_{-}^{\mu}(q_U^{\nu}q_+^{\rho}-g^{\nu\rho}(q_U\cdot q_+))+g^{\mu\nu}(q_{-}^{\rho}(q_U\cdot q_+) -q_+^{\rho}(q_U\cdot q_{-})) \\ +q_U^{\mu}(g^{\nu\rho}(q_{-}\cdot q_+)-q_+^{\nu}q_{-}^{\rho}) +g^{\mu\rho} 
(q_+^{\nu}(q_U\cdot q_{-}) -q_U^{\nu}(q_{-}\cdot q_+))).  \end{flalign*}
Again, due to the gauge structure, the amplitude with four longitudinal bosons doesn't play the major role, this amplitude goes with $k^{2d_{{\cal U}}-2}$.
Since $d_{{\cal U}}>1$ this alone would lead to unitarity violation. We come to that point later. But exactly as it was for the scalar unparticle, the leading contribution comes from the amplitudes with all four bosons being transversely polarised. This amplitude goes with $k^{2d_{{\cal U}}+2}$. If we again set $d_{{\cal U}}=1$ we also end up with a behaviour as in the Standard Model with the difference that in the Standard Model this occurs for longitudinal bosons only.\\ \\
Any amplitude can be written as a sum over partial waves. According the expansion in \cite{Jacob:1959at} we write our amplitude as\\
\begin{equation}\label{partial} \bra{\theta \phi \lambda_c \lambda_d}T(E)\ket{00\lambda_a\lambda_b}=16\pi \sum_{J}(2J+1)\bra{\lambda_c \lambda_d}T^J(E)\ket{\lambda_a \lambda_b}e^{i(\lambda-\mu)}d^J_{\lambda \mu}(\theta). \end{equation} \\
In this notation the $\lambda_i$ denote the helicity states of the particles, $\lambda=\lambda_a-\lambda_b$ and $\mu=\lambda_c-\lambda_d$. In the  case $\lambda=\mu=0$ the $d^j$-functions simplify to the Legendre-polynomials: $d^l_{00}(\theta)=P_L(\cos(\theta))$. $\bra{\theta \phi \lambda_c \lambda_d}T(E)\ket{00\lambda_a\lambda_b}$ is an ordinary T-matrix element.\\
To ensure unitarity\\
\begin{equation} |T^J|\le 1 \end{equation}
must be fillfulled and this relation has to be fullfilled for each partial wave. We assume that the strongest bounds come from the partial waves with small angular momentum. So we project our amplitude on the partial wave with $J=0$. This partial wave is given by\\
\begin{equation} P_0=\int_0^{\pi} {\cal M} \sin(\theta) d \theta. \end{equation} \\
For a scalar unparticle we have three different contributions to the leading amplitude. The first comes from the case when all polarization vectors are in the same state, which we denote as ${\cal M}_{(++++)}$. The first and the second sign denote the incoming $W^+$ and $W^{-}$, the third and the fourth the outgoing $W^+$ and $W^{-}$. The second and the third contribution come from the case when two particles have the same polarisation state. We denote them as ${\cal M}_{(++--)}$ and ${\cal M}_{(+-+-)}$. The three combinations $----$, $--++$ and $-+-+$ are related by symmetry and give the same results. All other amplitudes vanish.\\
We define $C_{WW,S}\equiv \frac{\lambda_1}{\Lambda^d_{\cal U}}$ as the coupling constant for a scalar unparticle to $W$-bosons.
For the leading term of the amplitudes for a scalar unparticle we then find\\
\begin{flalign}\label{Ms}\begin{split} {\cal M}_{S,++++}=&\frac{2^{d_{\cal U}+1} A_{d_{\cal U}} {C_{WW,S}}^2 e^{-i d_{\cal U} \pi } \left(e^{i d_{\cal U} \pi } (\cos (\theta )+1)^{d_{\cal U}}+2^{d_{\cal U}}\right)}{\sin(d_{\cal U}\pi)}k^{2d_{\cal U}}, \\
 {\cal M}_{S,++--}=&\frac{4^{d_{\cal U}+1} A_{d_{\cal U}} {C_{WW,S}}^2 e^{-i d_{\cal U} \pi }}{2\sin (d_{\cal U}\pi)}k^{2d_{\cal U}},\\
 {\cal M}_{S,+-+-}=&\frac{2^{d_{\cal U}+1} A_{d_{\cal U}} {C_{WW,S}}^2 (\cos (\theta )+1)^{d_{\cal U}}}{\sin(d_{\cal U}\pi)}k^{2d_{\cal U}}.\end{split}\end{flalign}\\
The second amplitude does not contribute to the $0$-th partial wave. Projections of the other two amplitudes lead to\\
\begin{flalign}\label{p0s}\begin{split} {\cal P}_{0,S,++++}=&\frac{2^{2 d_{\cal U}+2} A_{d_{\cal U}} {C_{WW,S}}^2 e^{-i d_{\cal U} \pi } \left(d_{\cal U}+e^{i d_{\cal U} \pi }+1\right)}{\sin(d_{\cal U}\pi)(d_{\cal U}+1)}k^{2d_{\cal U}},\\
{\cal P}_{0,S,+-+-}=&\frac{2^{2 d_{\cal U}+2} A_{d_{\cal U}} {C_{WW,S}}^2}{\sin(d_{\cal U}\pi)(d_{\cal U}+1)}k^{2d_{\cal U}}.
\end{split}\end{flalign}\\
Inserting (\ref{p0s}) into (\ref{partial}) and solving this equation for $C_{WW,S}$ leads to an upper bound on $C_{WW,S}$, dependent on the center-of-mass energy and on the scale dimension $d_{\cal U}$. This is illustrated in figure 3 for a center-of-mass energy of $1$ TeV and the different amplitudes. The allowed value for the coupling constant gets maximal when $d_{\cal U} \approx 1.8$.\\ \\
\psfrag{dU}{$d_{\cal U}$}
\psfrag{CWW}{$C_{WW,S}$ $[TeV^{-d_{\cal U}}]$}
\psfrag{CWWV}{$C_{WW,V}$ $[TeV^{-d_{\cal U}-1}]$}
\psfrag{s}{$\sqrt{s}$}
\begin{figure}
\parbox{16cm}{\includegraphics{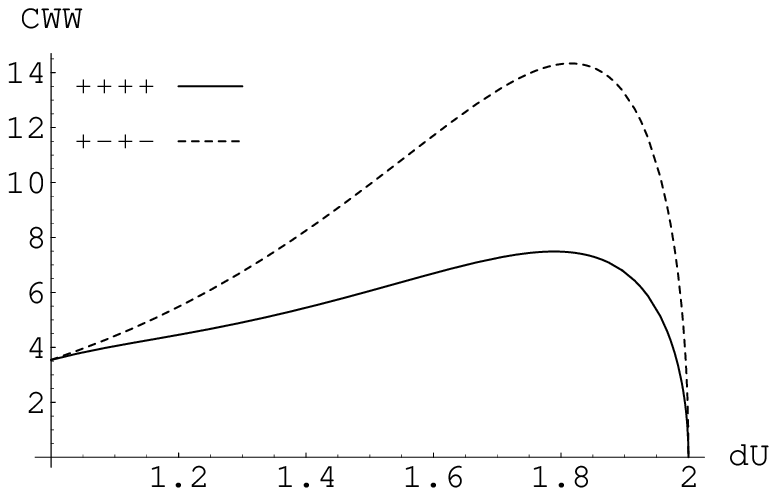}\includegraphics{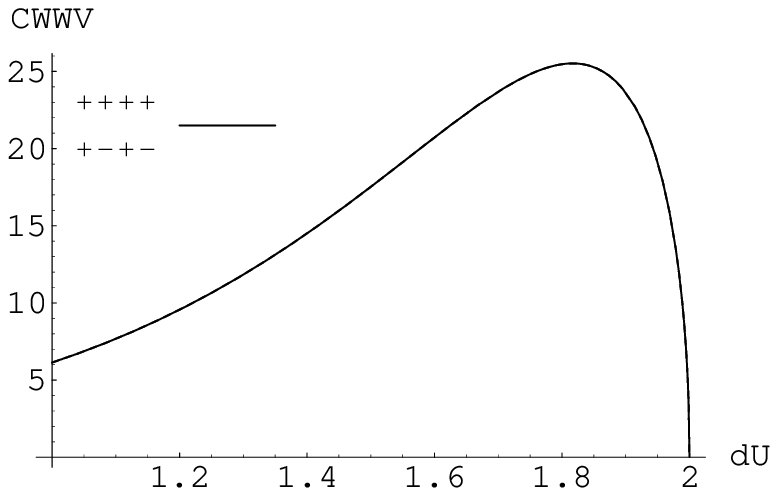}
\caption{Unitarity bounds on the coupling constant for a scalar- (left side) and a vector (right side) unparticle with $\sqrt{s}=1$ TeV and the different amplitudes contributing to the leading term.}}
\end{figure}
We repeat that procedure for a vector-like unparticle and define $C_{WW,V} \equiv \frac{\lambda_2}{\Lambda^{d_{\cal U}+1}}$. Again there are just three amplitudes that contribute to the leading term. They are again the three amplitudes ${\cal M}_{(++++)}$,${\cal M}_{(++--)}$ and ${\cal M}_{(+-+-)}$.\\ They are given by\\
\psfrag{dsigma}{$\frac{1}{\sigma}\frac{d\sigma}{d\theta}$}
\psfrag{theta}{$\theta$}
\psfrag{du}{$d_{\cal U}$}
\begin{figure}[h]
\begin{center}
\parbox{16cm}{\includegraphics{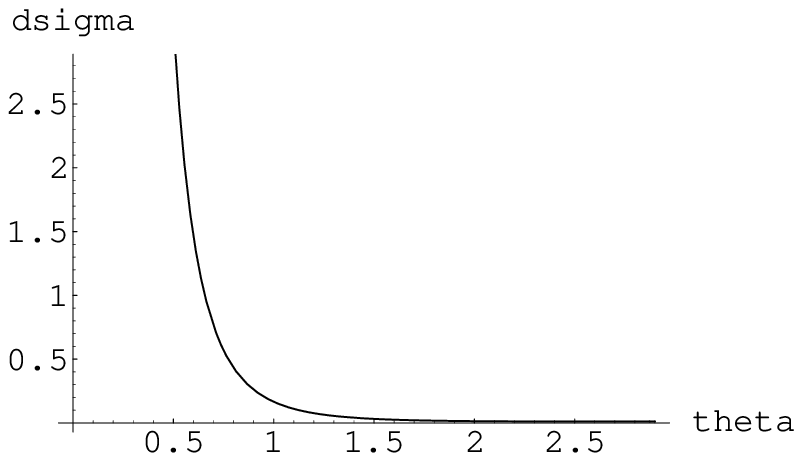} \includegraphics{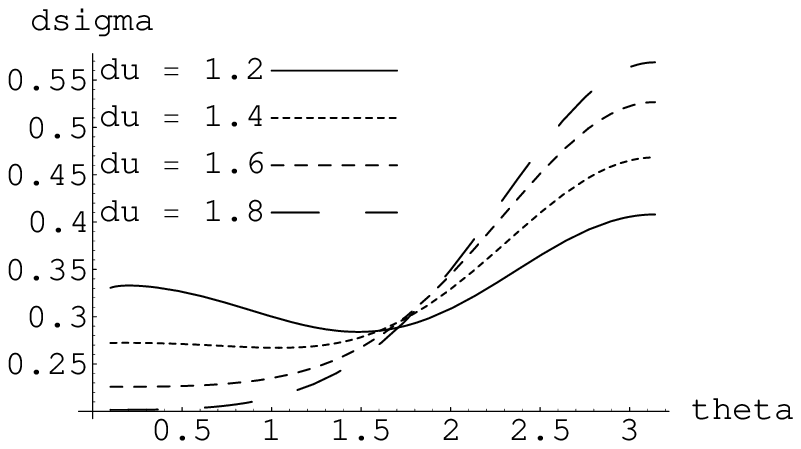}}\\
\vspace{0.8cm}
\parbox{5cm}{\includegraphics{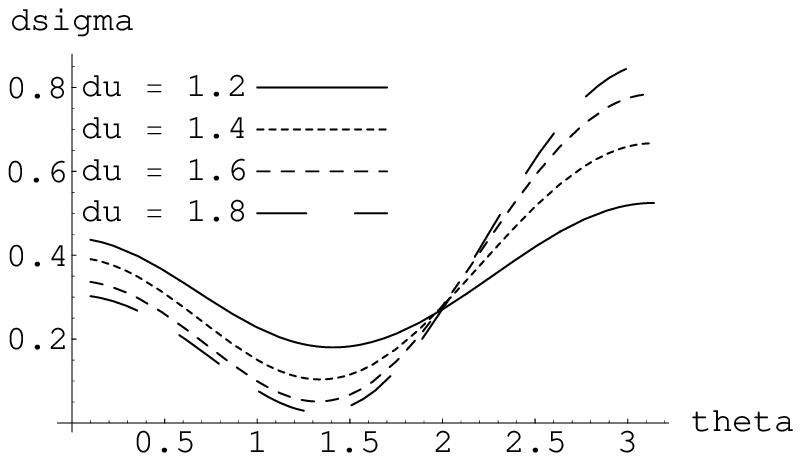}}
\caption{Normalised differential cross section $\frac{1}{\sigma}\frac{d\sigma}{d\theta}$ for the Standard Model, scalar and vector unparticles, for the leading terms in the high energy limit. We choose $\sqrt{s}=1$ TeV.}
\end{center}
\end{figure}
\begin{flalign}\label{mv} \begin{split} &{\cal M}_{V,++++}=\\ &-2^{d_{\cal U}}A_{d_{\cal U}}{C_{WW,V}}^2\frac{ \left(-\cos (\theta ) (\cos (\theta )+1)^{d_{\cal U}}+3 (\cos (\theta )+1)^{d_{\cal U}}+(-1)^{d_{\cal U}} 2^{d_{\cal U}+1} \cos (\theta )\right)}{\sin(d_{\cal U}\pi)}\\
& \hspace{15cm} \cdot k^{2d_{{\cal U}}+2},\\ \\
&{\cal M}_{V,++--}=\frac{-4^{d_{\cal U}+1} A_{d_{\cal U}} {C_{WW,V}}^2 e^{-i d_{\cal U} \pi } \cos (\theta )}{2\sin(d_{\cal U}\pi)}k^{2d_{{\cal U}}+2},\\
&{\cal M}_{V,+-+-}=\frac{2^{d_{\cal U}} A_{d_{\cal U}} {C_{WW,V}}^2 (\cos (\theta )-3) (\cos (\theta )+1)^{d_{\cal U}}}{\sin(d_{\cal U}\pi)}k^{2d_{{\cal U}}+2}.\\
\end{split} \end{flalign}
Concerning the partial waves, the projections of the first and the third amplitude are identical, the second amplitude gives no contribution to the $0$-th partial wave.\\ \\ \\ \\We find\\
\begin{flalign}\label{p0v}\begin{split}  P_{0,V,++++}=&P_{0,V,+-+-}=\frac{i2^{2 d_{\cal U}+2} A_{d_{\cal U}} {C_{WW,V}}^2 (d_{\cal U}+3)}{\sin(d_{\cal U}\pi)(d_{\cal U}^2+3 d_{\cal U}+2)}k^{2d_{{\cal U}}+2}.\\
 \end{split}\end{flalign}\\
From that we also get an upper bound on the coupling to vector unparticles. This is also shown in figure 3. The constraint on the vector unparticle is weaker than on the scalar unparticle but they are about the same order of magnitude.\\ \\
For unparticles the propagator depends on the scale dimension $d_{\cal U}$. That is why we expect a dependence of the angular distribution on $d_{\cal U}$. So the question arises if we can determine the scale dimension by measuring the angular distribution of the $W$-bosons.\\
On the upper left picture in figure 4 we plotted the differential cross section $\frac{1}{\sigma}\frac{d\sigma}{d \theta}$ of the Standard Model. It is normalised on the total cross section, i.e. the differential cross section for the Standard Model contribution is normalised on the total cross section of only the Standard Model. The differential cross section for the unparticles is normalised on the total cross section only including uparticles. On the upper right picture we plotted the differential cross section for a scalar unparticle for several values of $d_{\cal U}$ for a center-of-mass energy of $1$ TeV. On the lower picture the same is done for a vector unparticle. The distributions look very different compared to the Standard Model. This could make it possible to detect them at the LHC in VBF-processes for example \cite{Hankele:2006ma,Accomando:2005xp,Zhang:2003it,Gonzalez-Garcia:1999fq}.\\
Another interesting question is how we can distinguish a scalar unparticle from a vector unparticle. In the case of transverse polarisation, the distributions are too similar to be able to determine the lorentz structure. But there are differences in the pure longitudinal case. As stated earlier, the leading term for a scalar unparticle is proportional to $k^{2d_{\cal U}-4}$. So the Standard Model will dominate the cross section and the unparticle is negligible. The leading contribution for a vector unparticle in the longitudinal case is proportional to $k^{2d_{\cal U}-2}$. There we get an enhancement of the cross section that also may cause unitarity violation. In figure 5 we show the differential cross section for that case. We see that the distribution differs from that of the Standard Model and in that channel such a structure can only be caused by a vector unparticle.\\ \\ \\ 
\psfrag{dsigma}{$\frac{1}{\sigma}\frac{d\sigma}{d\theta}$}
\psfrag{theta}{$\theta$}
\psfrag{du}{$d_{\cal U}$}
\begin{figure}
\begin{center}
\parbox{16cm}{\includegraphics{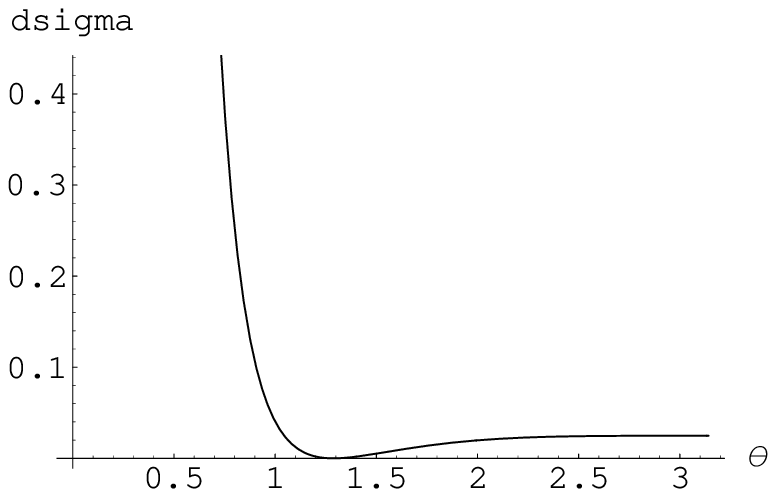} \includegraphics{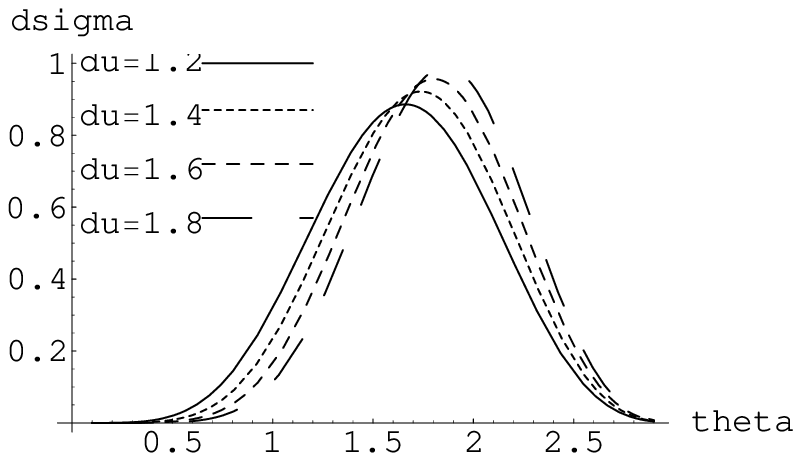}}
\caption{Normalised differential cross section $\frac{1}{\sigma}\frac{d\sigma}{d\theta}$ for the Standard Model and a vector unparticle when all $W$-bosons are longitudinally polarised.}
\end{center}
\end{figure}
\Large \bfseries 3 \hspace{0.5cm} Conclusions\\ \\ \mdseries
Starting from the fact that higher order dimensional operators receive strong constraints from unitarity arguments (see for instance \cite{Gounaris:1994cm}) and that the strongest bounds in the bosonic sector may come from the scattering of $W$-bosons, we calculated unitarity bounds for unparticles in this reaction. We found that the strongest bounds come from purely transverse polarised states and not from the longitudinal states due to the specific structure of the effective operators. Because of the dependence of the unparticle propagator on the scale dimension $d_{\cal U}$ we looked at the angular distribution of the outgoing $W$-bosons and found significant differences to the Standard Model for both a scalar and a vector unparticle. If we go to longitudinal polarisation it turned out that only a vector unparticle will contribute here in a significant manner so that in that case one could distinguish a scalar unparticle from a vector unparticle. So the whole process would in principle allow us to determine the spin characteristics of unparticles and may even allow to measure $d_{\cal U}$. \\ \\ \\
\Large \bfseries Acknowledgements\\ \\
\mdseries
We are very grateful to Thomas Gehrmann for helpful discussions and encouragement. We would also like to thank Ayres Freitas, Uli Haisch, Pedro Schwaller and Nurhana Tajuddin for useful discussions and comments. This work was supported by the Swiss National Science Foundation (SNF) under contract 200020-109162. 

\bibliographystyle{utphys}
\bibliography{literatur}

\providecommand{\href}[2]{#2}\begingroup\raggedright\begin{thebibliography}{10}

\bibitem{Georgi:2007ek}
H.~Georgi
\href{http://arXiv.org/abs/hep-ph/0703260}{{\tt hep-ph/0703260}}.

\bibitem{Georgi:2007si}
H.~Georgi
\href{http://arXiv.org/abs/arXiv:0704.2457 [hep-ph]}{{\tt arXiv:0704.2457
  [hep-ph]}}.

\bibitem{Banks:1981nn}
T.~Banks and A.~Zaks {\em Nucl. Phys.} {\bf B196} (1982)
189.

\bibitem{Cheung:2007ue}
K.~Cheung, W.-Y. Keung, and T.-C. Yuan
\href{http://arXiv.org/abs/arXiv:0704.2588 [hep-ph]}{{\tt arXiv:0704.2588
  [hep-ph]}}.

\bibitem{Li:2007by}
X.-Q. Li and Z.-T. Wei
\href{http://arXiv.org/abs/arXiv:0705.1821 [hep-ph]}{{\tt arXiv:0705.1821
  [hep-ph]}}.

\bibitem{Lu:2007mx}
C.-D. Lu, W.~Wang, and Y.-M. Wang
\href{http://arXiv.org/abs/arXiv:0705.2909 [hep-ph]}{{\tt arXiv:0705.2909
  [hep-ph]}}.

\bibitem{Duraisamy:2007aw}
M.~Duraisamy
\href{http://arXiv.org/abs/arXiv:0705.2622 [hep-ph]}{{\tt arXiv:0705.2622
  [hep-ph]}}.

\bibitem{Aliev:2007qw}
T.~M. Aliev, A.~S. Cornell, and N.~Gaur
\href{http://arXiv.org/abs/arXiv:0705.1326 [hep-ph]}{{\tt arXiv:0705.1326
  [hep-ph]}}.

\bibitem{Ding:2007bm}
G.-J. Ding and M.-L. Yan
\href{http://arXiv.org/abs/arXiv:0705.0794 [hep-ph]}{{\tt arXiv:0705.0794
  [hep-ph]}}.

\bibitem{Luo:2007bq}
M.~Luo and G.~Zhu
\href{http://arXiv.org/abs/arXiv:0704.3532 [hep-ph]}{{\tt arXiv:0704.3532
  [hep-ph]}}.

\bibitem{Liao:2007bx}
Y.~Liao
\href{http://arXiv.org/abs/arXiv:0705.0837 [hep-ph]}{{\tt arXiv:0705.0837
  [hep-ph]}}.

\bibitem{Chen:2007vv}
C.-H. Chen and C.-Q. Geng
\href{http://arXiv.org/abs/arXiv:0705.0689 [hep-ph]}{{\tt arXiv:0705.0689
  [hep-ph]}}.

\bibitem{Stephanov:2007ry}
M.~A. Stephanov
\href{http://arXiv.org/abs/arXiv:0705.3049 [hep-ph]}{{\tt arXiv:0705.3049
  [hep-ph]}}.

\bibitem{Fox:2007sy}
P.~J. Fox, A.~Rajaraman, and Y.~Shirman
\href{http://arXiv.org/abs/arXiv:0705.3092 [hep-ph]}{{\tt arXiv:0705.3092
  [hep-ph]}}.

\bibitem{Gounaris:1994cm}
G.~J. Gounaris, J.~Layssac, J.~E. Paschalis, and F.~M. Renard {\em Z. Phys.}
  {\bf C66} (1995) 619--632,
\href{http://arXiv.org/abs/hep-ph/9409260}{{\tt hep-ph/9409260}}.

\bibitem{Jacob:1959at}
M.~Jacob and G.~C. Wick {\em Ann. Phys.} {\bf 7} (1959)
404--428.

\bibitem{Hankele:2006ma}
V.~Hankele, G.~Klamke, D.~Zeppenfeld, and T.~Figy {\em Phys. Rev.} {\bf D74}
  (2006) 095001,
\href{http://arXiv.org/abs/hep-ph/0609075}{{\tt hep-ph/0609075}}.

\bibitem{Accomando:2005xp}
E.~Accomando and A.~Kaiser {\em Phys. Rev.} {\bf D73} (2006) 093006,
\href{http://arXiv.org/abs/hep-ph/0511088}{{\tt hep-ph/0511088}}.

\bibitem{Zhang:2003it}
B.~Zhang, Y.-P. Kuang, H.-J. He, and C.~P. Yuan {\em Phys. Rev.} {\bf D67}
  (2003) 114024,
\href{http://arXiv.org/abs/hep-ph/0303048}{{\tt hep-ph/0303048}}.

\bibitem{Gonzalez-Garcia:1999fq}
M.~C. Gonzalez-Garcia {\em Int. J. Mod. Phys.} {\bf A14} (1999) 3121--3156,
\href{http://arXiv.org/abs/hep-ph/9902321}{{\tt hep-ph/9902321}}.

\end{thebibliography}\endgroup
\end{document}